\documentclass[11pt]{article}

\author{Hao Chen\\
Software Engineering Institute\\
East China Normal University\\
Shanghai 200062, P.R. China\\
E-mail: haochen@sei.ecnu.edu.cn}
\title{\bf Restricted Parameter Range Promise  Set Cover Problems Are Easy}
\date{September, 2010}

\begin{document}

\maketitle
\begin{abstract}

Let $({\bf U},{\bf S},d)$ be an instance of Set Cover Problem, where
${\bf U}=\{u_1,...,u_n\}$ is a $n$ element ground set, ${\bf
S}=\{S_1,...,S_m\}$ is a set of $m$ subsets of ${\bf U}$ satisfying
$\bigcup_{i=1}^m S_i={\bf U}$ and $d$ is a positive integer. In STOC
1993 M. Bellare, S. Goldwasser, C. Lund and A. Russell proved the
NP-hardness to distinguish the following two cases of ${\bf
GapSetCover_{\eta}}$ for any constant $\eta > 1$. The Yes case is
the instance for which there is an exact cover of size $d$ and the
No case is the instance for which any cover of ${\bf U}$ from ${\bf
S}$ has size at least $\eta d$. This was improved by R. Raz and S.
Safra in STOC 1997 about the NP-hardness for ${\bf
GapSetCover}_{clogm}$ for some constant $c$. In this paper we prove
that  restricted parameter range subproblem is easy. For any given
function of $n$ satisfying $\eta(n) \geq 1$, we give a polynomial
time algorithm not depending on $\eta(n)$ to distinguish between\\

{\bf YES:} The instance $({\bf U},{\bf S}, d)$ where $d>\frac{2
|{\bf S}|}{3\eta(n)-1}$, for which there exists an exact cover of
size at most  $d$;\\

{\bf NO:} The instance $({\bf U},{\bf S}, d)$ where  $d>\frac{2
|{\bf S}|}{3\eta(n)-1}$, for which any cover from ${\bf S}$ has size
larger than
 $\eta(n) d$.\\

Thus the large part subproblem of the NP-hard promise set cover
problem is actually easy. The polynomial reduction of this
restricted parameter range set cover problem is constructed by using the lattice.\\

\end{abstract}

\section{Introduction}

{\bf Set Cover Problem} is a classical combinatorial optimization
problem. The instance of the problem is $({\bf U},{\bf S})$, where
where ${\bf U}=\{u_1,...,u_n\}$ is a $n$ element ground set, ${\bf
S}=\{S_1,...,S_m\}$ is a set of $m$ subsets. The goal is to find the
minimal size ${\bf S'} \subseteq {\bf S}$ such that $\bigcup_{S_i
\in {\bf S'}} S_i={\bf U}$. The classical result says that there is
the polynomial time algorithm approximating the optimal solution to
a factor $t$, where $t =max_i|\{j: u_i \in S_j\}|$ is the largest
number of the subsets in ${\bf S}$ for which one element in ${\bf
U}$ may belong(see [10, 12, 15]). The greedy algorithm with
polynomial time can also be used for approximating the optimal
solution to a $H_k=1+\frac{1}{2}+ \cdots +\frac{1}{k} \leq 1+ln k$
factor(see [5]), where $k=max_{i=1,...,m}\{|S_i|\}$ is the size of the largest set.\\

On the other hand, from the NP-completeness of $3$-dimensional
matching problem it is well-known that the following exact covering
problem by $3$-sets  is NP-complete: given an instance $({\bf
U},{\bf S})$, where ${\bf U}$ is a ground set of $3n$ elements and
${\bf S}$ is a collection of subsets  with $3$ elements, the goal is
to determine if there is a sub-collection ${\bf S'} \subseteq {\bf
S}$ of {\em pairwise disjoint} subsets(an exact cover) such that
$\bigcup_{S \in {\bf S'}}S={\bf U}$. In STOC 1993, Bellare,
Goldwasser, Lund and Russell proved that the promise problem of
approximating set cover to any constant factor is NP hard in [4](see
also [16]). Explicitly they proved that for any positive constant
$\eta > 1$ it is NP-hard to distinguish between the following {\bf
YES} and {\bf NO} instances. The YES instance is the instance for
which there is an exact cover of size $d$ , that is, there exist
pairwise disjoint subsets $S_{i_1},...,S_{i_d}$ in ${\bf S}$
satisfying $\bigcup_{j=1}^d S_{i_j}={\bf U}$,  and the No instance
is the instance for which any cover of ${\bf U}$ from ${\bf S}$ has
size at least $\eta d$. The result of R. Raz and S. Safra in  STOC
1997 ([19]) implies the NP-hardness of the promise problem ${\bf
GapSetCover}_{clogm}$ for some constant $c$. Feige [9] proved that
there cannot be a $(1-\varepsilon) ln m$ approximate algorithm for
the original set cover problem, for any $\varepsilon > 0$, unless
${\bf NP} \subseteq {\bf QP}$. Trevisan [20] indicated that Feige¡¯s
proof also implies that there is a constant c such that the {\em Set
Cover problem with sets of size $k$ }(where k is constant) has no
$(ln k- c ln ln k)$-approximate algorithm for the original set cover problem unless ${\bf NP}={\bf P}$. \\

A variant of {\bf Set Cover Problem} is the following vertex cover
problem for $k$-uniform hypergraphs. An edge in a hypergraph is a
subset of the vertices. A $k$-uniform hypergraph is ${\bf G}=({\bf
V},{\bf E})$, where ${\bf V}$ is the set of $n$ vertices and the
${\bf E}$ is set of edges and each edge in ${\bf E}$ is a $k$
element subset of ${\bf V}$. The {\bf Vertex Cover Problem} for
$k$-uniform hypergraphs($k$ is a constant) is defined as follows.
For any given $k$-uniform hypergraph, to find the minimum size
subset of vertices $V' \subseteq {\bf V}$ such that $V' \bigcap e
\neq \emptyset$ for each edge $e$. When $k=2$ it is the classical
vertex cover problem. There is a polynomial time greedy algorithm
approximating the {\bf Vertex Cover Problem} for $k$-uniform
hypergraphs to a factor $k$. Approximating the vertex cover problem
witnin a factor $k-1-\varepsilon$ for any $ \varepsilon>0$ and $k
\geq 3$ was proved NP-hard in [7]. Khot and Regev proved that
approximating the vertex cover problem for $k$-uniform hypergraphs
to the factor $k-\varepsilon$ for any $\varepsilon>0$ is NP-hard
under the
assumption that the {\bf Unique Game Conjecture} is true(see  [13]).\\

The main results of this paper are the following theorems.\\

{\bf Theorem 1.} {\em For any function of $n=|{\bf U}|$ satisfying
$\eta(n)
> 1$ there is a
polynomial time algorithm not depending on $\eta(n)$ to distinguish between\\

{\bf YES:} The set cover problem instance $({\bf U},{\bf S}, d)$
where $d>\frac{2|{\bf S}|}{3\eta(n)-1}$, for which there exists an
exact cover of size at most $d$;\\

{\bf NO:} The set cover problem instance $({\bf U},{\bf S}, d)$
where $d>\frac{2|{\bf S}|}{3\eta(n)-1}$,  for which any
cover from ${\bf S}$ has size larger than $\eta(n) d$.}\\

{\bf Corollary 1.} {\em  For any function of $m=|{\bf E}|$
satisfying $\eta(m) \geq 1$ there is a
polynomial time  algorithm not depending on $\eta(m)$ to distinguish between\\

{\bf YES}: The vertex cover problem for $k$-uniform hypergraphs
instance $({\bf G}=({\bf V},{\bf E}), d)$ satisfying $d>\frac{|{\bf
V}|}{2\eta(m)}$, for which there exists a
 subset $V'$ (exact vertex cover) of size at most $d$ such that $|V'\bigcap e|=1$ for each edge $e$ in ${\bf E}$;\\

{\bf NO}: The vertex cover problem for $k$-uniform hypergraphs
instance $({\bf G}=({\bf V},{\bf E}), d)$ satisfying $d>\frac{|{\bf
V}|}{2\eta(m)}$, for which any vertex cover has size larger than
 $\eta(m) d$.}\\

The reduction in Theorem 1 is based on  polynomial time solvable
computational problems for lattices.\\

\section{Proofs of the Main Results}

{\bf Proof of Corollary 1}. Set $|{\bf V}|=n$ and $|{\bf E}|=m$. Let
${\bf B}$ be the $m \times n$ matrix whose rows and columns
correspond to the the elements in ${\bf E}$ and ${\bf G}$
respectively. The entry $b_{ev}$ of ${\bf B}$ at the position
corresponding to the edge $e$ and vertex  $v$ is 1 if the vertex $v$
is in the edge $e$ and otherwise $b_{ev}=0$. Consider the lattice in
${\bf Z}^n$ defined by ${\bf L(B)}=\{{\bf x} \in
{\bf Z}^n: {\bf B} \cdot {\bf x}=0 \}$.\\

We prove the following two lemmas.\\

{\bf Lemma 1.}{\em 1) For {\bf YES} instance of the vertex cover
problem for $k$-uniform hypergraphs in Theorem 1, there exists a
vector in ${\bf L(B)}$ with  $n$ non-zero coordinates.}\\

{\em 2) For {\bf NO} instance of the vertex cover problem for
$k$-uniform
hypergraphs in Theorem 1, any vector in ${\bf L(B)}$ has  at most $2(n-\eta(m) d)$ non-zero coordinates.}\\

{\bf Proof.} Let ${\bf 1}$ be the vector in ${\bf Z}^n$ (or ${\bf
Z}^m$) with all $n$(or $m$) coordinates $1$. For any instance of
vertex cover problem for $k$-uniform hypergraphs, it is obvious
${\bf B} \cdot {\bf 1}=k{\bf 1}$. For {\bf Yes} case, there exists a
integral vector ${\bf x} \in {\bf Z}^n$ with at most $d$ non-zero
coordinates which equals to $1$ such that ${\bf B} \cdot {\bf
x}={\bf 1}$, since there exists an exact vertex cover of size at
most $d$. Thus ${\bf B} \cdot (k{\bf x}-{\bf 1})=0$. It is obvious
that $(k{\bf x}-{\bf 1})$ has $n$ non-zero coordinates.
The conclusion in 1) is proved.\\

For {\bf NO} case, let ${\bf x}={\bf x}^{+}-{\bf x}^{-} \in {\bf
Z}^n$ be any vector in the lattice ${\bf L(B)}$, where ${\bf x}^{+}$
and ${\bf x}^{-}$ be two vectors in ${\bf Z}^n$ with all their
coordinates non-negative integers. Set ${\bf x'} \in {\bf Z}^n$  an
integral vector which equals to ${\bf x}^{+}$ at the non-zero
positions of ${\bf x}^{+}$ and takes any positive integer at the
zero  positions of ${\bf x}^{+}$. It is obvious every coordinate of
the vector ${\bf B} \cdot {\bf x'}$ is a positive vector, since this
vector is a linear combination of all columns of the matrix ${\bf
B}$ with positive coefficients. Then we have every coordinates of
the integral vector ${\bf B} \cdot ({\bf x'}-{\bf x}^{+})$ is
positive integer. Note that ${\bf x'}-{\bf x}^{+}$ has all
coordinates non-negative integers. Thus the vertex corresponding to
the non-zero positions of the vector ${\bf x'}-{\bf x}^{+}$ is a
vertex cover. Thus we have $n-|supp({\bf x}^{+})| \geq \eta(m) d$,
where $supp({\bf v}) \subseteq [n]$ is the set of non-zero positions
of the vector ${\bf v }\in {\bf Z}^n$. Similarly we can
prove that $|supp({\bf x}^{-})| \leq n- \eta(m) d$. The conclusion is proved.\\

 {\bf Lemma 2.} {\em For {\bf NO} instance of the vertex
cover problem for $k$-uniform hypergraphs in Theorem 1, there exists
a subspace ${\bf R}^{2(n-\eta(m) d)}\times \{0\}^{2\eta(m) d-n}$ in
${\bf Z}^n \otimes {\bf R}$
such that ${\bf L(B)} \subseteq {\bf R}^{2(n-\eta(m) d)}\times \{0\}^{2\eta(m) d-n}$.}\\

{\bf Proof.} Let ${\bf x}=(x_1,...,x_n)$ be a vector in the lattice
${\bf L(B)}$ with the largest number of non-zero
coordinates$|supp({\bf x})|$. From 2) of Lemma 1, we have
$|supp({\bf x})| \leq 2(n-\eta(m) d)$. Suppose there exists a vector
${\bf y}=(y_1,....,y_n) \in {\bf L(B)}$ which has one non-zero
coordinate position outside $supp({\bf x})$. It is clear there
exists a integer $t$ such that $x_i+ty_i \neq 0$ for those indices
$i$ satisfying $x_i \neq 0$ or $y_i\neq 0$. Then the integral vector
${\bf x} +t{\bf y} \in {\bf L(B)}$ has larger support than ${\bf
x}$, which has at least $|supp({\bf x})|+1$ non-zero positions.
This is a contradiction.\\

Fixed any function  $\eta(m)  \geq 1$ we have the following algorithm.\\

{\bf Algorithm}\\

{\bf Input}: The vertex cover problem for $k$ uniform hypergraphs instance $({\bf G}=({\bf V},{\bf E}),d)$
satisfying
$d >\frac{|{\bf E}|}{2\eta(m)}$, for which the instance is {\bf YES} or {\bf NO} of Theorem 1.\\

{\bf Output}: {\bf YES} or {\bf NO}.\\

{\em Step 1.} Find a base of the lattice ${\bf L(B)}$;\\

{\em Step 2.} Check if there are at least $2\eta(m) d-n$ coordinate
positions such that all these base vectors in ${\bf Z}^n \otimes
{\bf R}$ are zero at the  $2\eta(m) d-n$ positions. If yes, answer
{\bf NO}, if for every coordinate position, there exists a base
vector which is non-zero at this position, answer {\bf YES}.  \\

 {\bf Lemma 3.} {\em The above algorithm answers the promise problem
 in Theorem 1 correctly in polynomial time.}\\

 {\bf Proof.} Since $2(n-\eta(m) d) < n$, it is clear the algorithm solve the promise problem
 correctly from Lemma 1 and 2. On the other hand the step 1 and 2
 are all in polynomial time in $mn$ bits(the inputs size of the
 matrix ${\bf B}$, see [18]). Thus the conclusion is proved.\\

{\bf Proof of Theorem 2.} Set $|{\bf U}|=n$ and $|{\bf S}|=m$. Let
${\bf B}$ be the $n \times m$ matrix whose rows and columns
correspond to the the elements in ${\bf U}$ and subsets in ${\bf S}$
respectively. The entry $b_{uS}$ of ${\bf B}$ at the position
corresponding to the point $u \in {\bf U}$ and subset $S \in {\bf
S}$ is 1 if the element $u$ is in the subset $S$ and otherwise
$b_{uS}=0$. Consider the lattice in ${\bf Z}^m$ defined by ${\bf
L(B)}=\{{\bf x} \in {\bf Z}^m: {\bf B} \cdot {\bf x}=0 \}$. Let
${\bf B'}=({\bf B},{\bf 1})$ be the $n \times (m+1)$ matrix with the
all $1$ vector ${\bf 1} \in {\bf Z}^n$ appending to the matrix ${\bf
B}$. We define lattice ${\bf L(B')}=\{{\bf x} \in {\bf Z}^{m+1}:{\bf
B'} \cdot {\bf x}=0\} \subseteq {\bf Z}^{m+1}$. \\

{\bf Lemma 4.} {\em  For {\bf NO} instance of the set cover problem
in Theorem 2, any vector in ${\bf L(B)}$ has at most $2(m-\eta(n)
d)$
non-zero coordinates.}\\

{\bf Proof.} Let $({\bf U},{\bf S}, d)$ be a {\bf NO} instance. Let
${\bf x}={\bf x}^{+}-{\bf x}^{-} \in {\bf Z}^m$ be any vector in the
lattice ${\bf L(B)}$, where ${\bf x}^{+}$ and ${\bf x}^{-}$ be two
vectors in ${\bf Z}^m$ with all their coordinates non-negative
integers. Set ${\bf x'} \in {\bf Z}^m$  an integral vector which
equals to ${\bf x}^{+}$ at the non-zero positions of ${\bf x}^{+}$
and takes any positive integer at the zero  positions of ${\bf
x}^{+}$. It is obvious every coordinate of the vector ${\bf B} \cdot
{\bf x'}$ is a positive vector, since this vector is a linear
combination of all columns of the matrix ${\bf B}$ with positive
coefficients. Then we have every coordinates of the integral vector
${\bf B} \cdot ({\bf x'}-{\bf x}^{+})$ is positive integer. Note
that ${\bf x'}-{\bf x}^{+}$ has all coordinates non-negative
integers. Thus the subsets in ${\bf S}$ corresponding to the
non-zero positions of the vector ${\bf x'}-{\bf x}^{+}$ is a cover
of the set ${\bf U}$. Thus we have $m-|supp({\bf x}^{+})| \geq
\eta(n) d$, where $supp({\bf v}) \subseteq [m]$ is the set of
non-zero positions of the vector ${\bf v }\in
{\bf Z}^m$. Similarly we can prove that $|supp({\bf x}^{-})| \leq m- \eta(n) d$. The conclusion is proved.\\

{\bf Proof of Corollary 1.} From the proof of Lemma 4 we note that
for the {\bf NO} instances of the promise set cover problem in
Corollary 1, the lattice ${\bf L(B)}$ is the zero lattice. Thus it
is easy to distinguish the {\bf YES} and {\bf NO} instances in
Corollary 1. We can check if the lattice ${\bf L(B)}$ is zero and
then find a rational solution ${\bf y}$ of the system of linear
equations ${\bf B} \cdot {\bf y}={\bf 1}$. When ${\bf L(B)}$ is not
zero, it is {\bf YES} instance. When ${\bf L(B)}=0$, it is {\bf YES}
instance if the Hamming weight of ${\bf y}$ is smaller than $m$ and
it is {\bf NO} instance if the Hamming weight of ${\bf y}$ is $m$.\\

Similarly as Lemma 2 we can prove the following lemma.\\

{\bf Lemma 5.} {\em For {\bf NO} instance of the set cover problem
in Theorem 2, there exists a subspace ${\bf R}^{2(m-\eta(n)
d)}\times \{0\}^{2\eta(n) d-m}$ in ${\bf Z}^m \otimes {\bf R}$ such
that ${\bf L(B)} \subseteq {\bf R}^{2(m-\eta(n) d)} \times
\{0\}^{2\eta(n) d-m}$.}\\

{\bf Lemma 6.} {\em For the {\bf YES} instance of  the set cover
problem in Theorem 2, if ${\bf L(B)} \subseteq {\bf R}^{2(m-\eta(n)
d)}\times \{0\}^{2\eta(n) d-m}$  for some subspace ${\bf
R}^{2(m-\eta(n))}\times \{0\}^{2\eta(n) d-m} \subseteq {\bf Z}^m
\otimes {\bf R}$. Set $P: {\bf Z}^m \otimes {\bf R} \rightarrow
\{0\}^{2(m-\eta(n) d)} \times {\bf R}^{2\eta(n) d-m}$ be the
projection to that fixed orthogonal complimentary space ${\bf
R}^{m-2 \eta(n) d}\times \{0\}^{2\eta(n) d-m}$. We have that ${\bf
L(B')}-{\bf L(B)}$ is not empty and any lattice vector ${\bf x}$ in
${\bf L(B')}-{\bf L(B)}$ satisfies
$||P({\bf x})|| \leq \sqrt{d}$.}\\

{\bf Proof.} For the {\bf YES} instance of Theorem 2, there exists a
${\bf x_0} \in \{0,1\}^m$ with at most $d$ $1$ coordinates
satisfying ${\bf B} \cdot {\bf x_0}={\bf 1}$ from the condition
there exists an exact cover of size at most $d$. Thus ${\bf
L(B')}-{\bf L(B)}$ is not empty  and any vector in ${\bf L(B')}-{\bf
L(B)}$ is of the form ${\bf x_0}+{\bf x}$ where ${\bf x} \in {\bf
L(B)} \subseteq {\bf R}^{2(m-\eta(n) d)}$. The projection $P$ 's
image is in the orthogonal complimentary of ${\bf R}^{2(m-\eta(n)
d)}$. From the property ${\bf x} \in {\bf R}^{2(m-\eta(n) d)}$, we
know that there are at most $d$ nonzero coordinates (which are $1$)
in $P({\bf x_0}+{\bf x})$. The conclusion is proved.\\

For any function satisfying $\eta(n) >1$ we have the following algorithm.\\

{\bf Algorithm}\\

{\bf Input}: The set cover problem  instance $({\bf U},{\bf S},d)$
satisfying $d >\frac{2|{\bf S}|}{3\eta(n)-1}$, for which the instance is {\bf YES} or {\bf NO} of Theorem 2.\\

{\bf Output}: {\bf YES} or {\bf NO}.\\

{\em Step 1.} Find a base of the lattice ${\bf L(B)}$;\\

{\em Step 2.} Check if there are at least $2\eta(n) d-m$ coordinate
positions  such that all these base vectors in ${\bf Z}^m \otimes
{\bf R}$ are zero at the  $2\eta(n) d-m$ positions. If there are
less
than $ 2\eta(n) d-m$ zero positions for all these base vectors ,  answer {\bf YES}.\\

{\em Step 3.} If the answer of the previous step is yes, check if
the lattice ${\bf L(B')}$ equals to ${\bf L(B)}$ by the natural
inclusion ${\bf L(B)} \rightarrow {\bf L(B')}$.
If they are the same, answer {\bf NO}.\\

{\em Step 4.} If the answer of the previous step is not. The ${\bf
L(B')}$ and ${\bf L(B)}$ are not the same.  Set $P: {\bf Z}^m
\otimes {\bf R} \rightarrow {\bf R}^{m- 2\eta(n) d}$ be the
projection to that fixed ${\bf R}^{m-2\eta(n) d}$. Take an arbitrary
lattice vector ${\bf x}$ in ${\bf L(B')}$ not in ${\bf L(B)}$, check
if
$||P(x)||>\sqrt{d}$, if yes answer {\bf NO}, if not, answer {\bf YES}.\\

{\bf Lemma 7.} {\em The above algorithm answers  the promise problem
 in Theorem 2 correctly in polynomial time.}\\

 {\bf Proof.} From the condition $d>\frac{2|{\bf S}|}{3\eta(n)-1}$, we
 have $\eta(n) d-2(m-\eta(n) d)>d$. If there exists a vector ${\bf x_0} \in
 {\bf L(B')}-{\bf L(B)}$ for {\bf NO} instance, then $P({\bf x_0}+{\bf x})
 >\sqrt{\eta(n)
 d-2(m-\eta(n) d)}>\sqrt{d}$. On the other hand, it is clear that all computation
 complexity in Step 1-4 is polynomial
 time of $mn$. The conclusion is proved.\\

\section{Conclusion}

The implications of Theorem 1 are as follows. Firstly  the $\frac{3
\eta-3}{3\eta-1}$ part subproblem of the NP-hard promise set cover
problem is easy when the function $\eta(n)$ is a constant. Secondly if a problem can be reduced
to the promise set cover problem within the restricted parameter
range as described in Theorem 1 within polynomial time, it is an
easy problem. Thirdly, from the results in Theorem 1, we can see
that the instances constructed from {\bf SAT} problem reduction to
the set cover problem(or vertex cover
problem for $k$-uniform hypergraph problem)
in the previous works [4, 16, 19, 7] are not in the restricted parameter range in this paper.\\

{\bf Acknowledgment.} The work was supported by
Grant 10871068 of NSFC and DNRF-NSFC joint grant 11061130539 .\\

\begin{center}
REFERENCES
\end{center}

[1] M. Ajtai, The shortest vector problem in $L_2$ is NP-hard for
randomized reductions(extended astract), Proc. STOC 1998, pages
10-19.\\

[2] S. Arora, L. Babai, J. Stern and Z. Sweedyk, The hardness of
approximating optima in lattices, codes and systems of linear
equations, Journal of Computer and System Science, 54(2), pages
317-331, 1997, Preliminary version, FOCS 1993.\\

[3] N. Bansal and S. Khot, Inapproximability of hypergraph vertex
cover and applications to scheduling problems, preprint 2009, see
http://cs.nyu.edu\\/~khot/publications.html.\\

[4] M. Bellare, S. Goldwasser, C. Lund and A. Russell, Efficient
multi-prover interactive proofs with applications to approximation
problems, Proc. STOC 1993, pages 113-131, 1993.\\

[5] V. Chvatal, A greedy heuristic for the set-covering problem.
Mathematics of Operations Research, Vol. 4, pages 233-235, 1979. \\

[6] I. Dinur, G. Kindler, R. Raz, and S. Safra, An improved lower
bound for approximating CVP. Combinatorica, Vol.23(2), pages
205-243, 2003, Preliminary version, STOC 1999.\\

[7] I. Dinur, V. Guruswami, S. Khot and O. Regev, A new multilayered
PCP and the hardness of hypergraph vertex cover, SIAM Journal on
Computing, Vol.34, no.5, 2005, pages 1129-1146, Preliminary version,
STOC 2003.\\

[8] I. Dumer, D. Micciancio and M. Sudan, Hardness of approximating
the minimum distance of a linear code, Proc FOCS 1999, Journal
version, IEEE Transactions on Information Theory, 49(1), pages
22-37, 2003.\\

[9] U. Feige, A threshold of $lnn$ for approximating set cover,
Journal of ACM, vol.45, no.4, pages 634-652, 1998.\\

[10] D. Hochbaum, Approximation algorithms for set covering and
vertex cover problems. SIAM Journal on Computing, vol. 11, pages
555-556, 1982.\\

[11] I. Haviv and O. Regev, Tensor-based hardness of the shortest
vector problem to within almost polynomial factors, Proc. STOC
2007, pages 469-477.\\

[12] D. S. Johnson, Approximation algorithms for combinatorial
problems. Journal of Computer and System Sciences, vol.9, pages
256-278, 1974.\\

[13]  S. Khot and O. Regev, Vertex cover might be hard to
approximate to within 2 ,  Proc. of the 18th IEEE Conference on
Computational Complexity, 2003. \\

[14] S. Khot, Hardness of approximating the shortest vector problem
in lattices, Journal ACM, 52(5), pages 789-808, 2005,
Preliminary version,  FOCS 2004.\\

[15] L. Lovasz, On the ratio of optimal integral and fractional
covers. Discrete Mathematics, Vol.13, pages 383-390, 1975. \\

[16] C. Lund and M. Yannakakis, Hardness of approximating
minimization problem, J. ACM, 41(5), pages 960-981, 1994,
Preliminary version, STOC 1993.\\

[17] D. Micciancio, The shortest vector problem is NP-hard to
approximate within some constant, SIAM Journal of Computing,
 30(6), pages 2008-2035, 2001, Preliminary version,  FOCS 1998.\\

[18] D.Micciancio and S. Goldwasser, Complexity of lattice problems,
A Cryptographic perspective, Kluwer Academic Publishers, 2002.\\

[19] R. Raz and S. Safra, A sub-constant error-probability low
degree test and a sub-constant error-probability PCP characteristic
of NP,
Proc. STOC 1997, pages 475-484.\\

[20] L. Trevisan, Non-approximability results for optimization
problems on bounded degree instances, Proc. STOC 2001, pages 453-461, 2001. \\

\end{document}